\documentclass[12pt,a4paper]{article}
\usepackage[dvips]{graphicx}
\begin{document}
\sloppypar
\bigskip

\title{
Non-Markovian collision integral in Fermi-systems
}

\author{V.A. Plujko, S.N. Ezhov, O.M. Gorbachenko, M.O.Kavatsyuk}

\date{}

\maketitle

\vspace{-1cm}
\begin{center}
Nuclear Physics Department,Physical Faculty,
Taras Shevchenko National University,\\
 Pr. Acad. Glushkova, 2, Bldg. 11, 03022 Kiev, Ukraine\\
E-mail: plujko@mail.univ.kiev.ua
\end{center}

\begin{center}
\parbox{0.8\textwidth}{\small
The non-Markovian collision integral is obtained
on the base of the Kadanoff-Baym equations for Green functions in a form
with allowance for small retardation effects.
The collisional relaxation times and damping width of the giant isovector
dipole resonances in nuclear matter are calculated. For an infinite Fermi
liquid the dependence of the relaxation times on the collective vibration
frequency and the temperature corresponds to the Landau's prescription.
}
\end{center}

\section{Introduction}
The dissipative properties of many-body systems, specifically
transport coefficients for viscosity as well as the damping of the
collective excitations, are strongly governed by the interparticle
collisions. In semi-classical approach these collisions are
described by collision integral in kinetic equation. It allows us
to involve a viscosity into motion equations similar to the
phenomenological Navier-Stockes ones \cite{AK58}-\cite{JCVL99}.
An advantage of such approach is a conceptual clarity and a possibility
to use many results from a general macroscopic physics.

For correct description of the collision relaxation rates in
systems with a fast variation of the mean field, it is necessary
to incorporate the memory (retardation) effects
\cite{RRG85}-\cite{mf99}. We investigate the form of the
non-Markovian linearized  collision integral with retardation
using the Kadanoff-Baym equations\cite{KB1962,D1986} for Green functions.
This method enables immediately to take into account self-consistent mean
field.

The non-Markovian collision term of the semiclassical linearized
Vlasov-Landau equation was obtained with the use of the
Kadanoff-Baym equations for the Green functions in
Refs.~\cite{kmp92,kp94}. The kinetic equation  was considered
in one-component Fermi liquid with the use of  the linear approximation
on deviation from overall equilibrium with a nonequilibrium distribution function
$\delta f = f(\vec{p}, \vec{r},t)- \bar{f}( \bar{\epsilon})
\propto \exp (-i\omega t)$, where $f(\vec{p}, \vec{r},t)$ is
an exact distribution function and $\bar{f}$ is an equilibrium
distribution function at equilibrium single-particle
energy $ \bar{\epsilon}$. An expression for linearized collision
integral was obtained in the Born approximation on two-body
collisions in the following general form:
\begin{equation}
J({\vec p}, {\vec r}, t) = J^{(1)}({\vec p}, {\vec r}, t) +
J^{(2)}({\vec p}, {\vec r}, t) ,
\label{eqJ}
\end{equation}
where the first component corresponds to variation of the
distribution function and second one is connected with variation
$\delta U = \epsilon(\vec{p}, \vec{r},t)-\bar{\epsilon}$ of mean field
with $\epsilon(\vec{p}, \vec{r},t)$ for actual single-particle energy.

  One can  see that this relationship does not agree with general
form of the  linearized Landau-Vlasov equation in Fermi liquid
in the limiting case of the states with week time-dependence of
nonequilibrium component of distribution function,  when
$\partial \delta f / \partial t \to 0$.  Indeed, the  linearized
Landau-Vlasov equation in quasi-homogeneous systems can be presented
in the following form:
\begin{equation}
\frac{ \partial \delta f}{\partial t}+ \frac{p}{m} ( \vec{\hat{p}}
\cdot \frac{\partial}{\partial {\vec r}} ) \delta \bar{f} = J ,
\label{LVone}
\end{equation}
where $\delta \bar{f}$ is a linear deviation of the exact distribution
function $f(\vec{p}, \vec{r},t)$ from the distribution function of the
equilibrium shape $\bar{f}(\epsilon)$ evaluated with actual single-particle
energy $ \epsilon = \bar{\epsilon}+ \delta U$,
\begin{equation}
\delta \bar{f}= \delta f -
\frac{d \bar{f}}{d \bar{\epsilon}} \delta U =
\delta f - \bar{f}(\epsilon) .
\label{flocal}
\end{equation}

The derivative $\partial \delta f / \partial t$ can be omitted in
Eq. (\ref{LVone}) for slightly time-dependent states and the
collision integral should be  a functional, $\Psi$, of the
$\delta\bar{f}$,
 \begin{equation}
J = \Psi (\delta\bar{f}) \equiv  \frac{p}{m} ( \vec {\hat{p}}
\cdot \frac{\partial}{\partial \vec{r}} ) \delta \bar{f},
\label{Intshape}
\end{equation}
in order to the  Landau-Vlasov equation would be fulfilled in this case.
The relationships (\ref{eqJ}) and (\ref{Intshape})
are generally in contradictions and derivation of the form of non-Markovian
collision integral from Refs.~\cite{kmp92,kp94} can be revised.

It was pointed in Refs. \cite{LifPit,PW1973} that collisional
integrals of the type (\ref{Intshape}), $ J \equiv
J(\delta\bar{f})$, are general form of the collision integrals
between quasiparticles  in Fermi liquid without retardation and
they lead to  a local equilibrium state described by  distribution
function $f_{l.e.}$ = $\bar{f}(\epsilon=\bar{\epsilon}+ \delta
U)$. Due to this we use in the following this terminology.

In this contribution an expression for non-Markovian collision
integral of the linearized Vlasov-Landau transport equation from
Refs. \cite{kmp92,kp94} is modified in Sect. 2 for the case of
slightly time-dependent states,i.e. in a form which allows for
reaching the local equilibrium in system.

In Sect.3 calculations of the relaxation times and damping width
of the collective vibration in nuclear matter with the use
collision integral with retardation effect are presented.

\section{Non-Markovian linearized  collision integral within
semiclassical approach}

In order to obtain linearized Vlasov-Landau equation with
collision integral we use the mixed $\{ \vec{p},\vec{r} \}$
Weyl-Wigner representation
for  single-particle Green functions (correlation functions)
$G^{>}(1,1')$, $G^{<}(1,1')$:
\begin{equation} 
\begin{array}{l} \vspace{1mm}
{\displaystyle f(\vec{p},\vec{r},t) \equiv
g^<(t,t;\vec{p},\vec{r})= \int d\vec{r}^{\,\,'}
\exp(-\frac{i}{\hbar}\vec{p}\vec{r}^{\,\,'}) g^<(t,t;\vec{r}+
\frac{\vec{r}^{\,\,'}}{2},\vec{r}-\frac{\vec{r}^{\,\,'}}{2});} \\
\vec{r}^{\,\,'}=\vec{r_1}-\vec{r}_{1'},
\vec{r}=\frac{1}{2}(\vec{r}_1-\vec{r}_{1'}) ,
\end{array}
\end{equation}
where $f(\vec{p},\vec{r},t)$ is one-body distribution function and
Green functions  are determined by\cite{KB1962,D1986}
\begin{equation}\label{eq5.1}
\begin{array}{l}
G^{>}(1,1^\prime ) = -i
\,\langle\Psi (1)\,\Psi^{+}(1^\prime )\rangle \,,\quad
t_1>t_1^{'}\,,\\  G^{<}(1,1^\prime ) = i
\,\langle\Psi^{+}(1^\prime )\,\Psi (1)\rangle  \,,\quad
t_1<t_1^{'}\,.
\end{array}
\end{equation}
Here, $\Psi^{+}(1)$ and $\Psi (1)$ are the operators of creation
and annihilation of a fermion; the symbol $1$ includes both space
and time variables, namely, $1 \equiv \{ \vec {r}_1, t \}$ (we omit
isotopic and spin variables); and the expectation values in
Eq.(\ref{eq5.1}) are calculated for the ground state or for the
ensemble of initial states if a temperature of system is not zero.

The Green functions $G^{>}(1,1^{'} )$  and $G^{<}(1,1^{'})$
satisfy the Kadanoff-Baym equation
\begin{equation}
\begin{array}{l}
{\displaystyle \left[i\hbar\frac{\partial}{\partial t_1} +
\frac{\hbar^2}{2m}\nabla^{2}_{1}-U(1)\right] G^{><}(1,1^{'}) =
I^{><}_1(1,1^{'}) - I^{><}_2(1,1^{'}),} \vspace{1mm}\\
{\displaystyle \left[-i\hbar\frac{\partial}{\partial t_1^{'}} +
\frac{\hbar^2}{2m} \nabla^{2}_{1^{'}}-U(1^{'})\right]
G^{><}(1,1^{'}) = J^{><}_1(1,1^{'}) -J^{><}_2(1,1^{'}) ,}
\end{array}
\end{equation}
where the single-particle potentials $U(1)$, $U(1^{'})$
are determined by the relations
\begin{equation}
\begin{array}{l}
{\displaystyle U(1)G^{><}(1,1^{'})=\int d\vec{r}_2
\Sigma_0(\vec{r}_1,\vec{r}_2,t_1) G^{><}(\vec{r}_2,t_1;1'),}
\vspace{1mm} \\ {\displaystyle U(1')G^{><}(1,1')=\int
d\vec{r}_2G^{><} (1;\vec{r}_2,t_{1'})
\Sigma_0(\vec{r}_2,\vec{r}_{1'},t_{1'}) }
\end{array}
\end{equation}
with
\begin{equation}
\begin{array}{rcl}
\Sigma_0(\vec{r}_1,\vec{r}_2,t)& =& {\displaystyle
-i\delta(\vec{r}_1-\vec{r}_2) \int
d\vec{r}v(|\vec{r}_1-\vec{r}|)G^{<}(\vec{r},t_1;\vec{r},t_1)}
\vspace{1mm} \\ & & +
iv(|\vec{r}_1-\vec{r}_2|)G^{<}(\vec{r}_1,t_1;\vec{r}_2,t_1),
\end{array}
\end{equation}
and $v(|\vec{r}_1-\vec{r}_2|)$ for two body potential.

The functions $I^{><}_{1,2}$ and $J^{><}_{1,2}$ are correlation
integrals of the form
\begin{equation}
\begin{array}{l}
{\displaystyle
I^{><}_1(1,1^{'})=\frac{1}{\hbar}\int_{t_0}^{t_1}dt_2
\left[\Sigma^{>}(1,2)-\Sigma^{<}(1,2)\right]G^{><}(2,1^{'}),}
\vspace{1mm} \\ {\displaystyle I^{><}_2(1,1^{'}) =
\frac{1}{\hbar}\int_{t_0}^{t_1^{'}} dt_2
\Sigma^{><}(1,2)\left[G^>(2,1^{'})-G^<(2,1^{'})\right],}
\vspace{1mm}\\
{\displaystyle
J^{><}_1(1,1^{'})=\frac{1}{\hbar}\int_{t_0}^{t_1}dt_2
\left[G^{>}(1,2)-G^{<}(1,2)\right]\Sigma^{><}(2,1^{'}),}
\vspace{1mm}
\\{\displaystyle
 J^{><}_2(1,1^{'})=\frac{1}{\hbar}\int_{t_0}^{t_1^{'}}dt_2
G^{><}(1,2)\left[\Sigma^{>}(2,1^{'})-\Sigma^{<}(2,1^{'})\right] .}
\end{array}
\end{equation}
These correlation integrals have the general form with allowance
for retardation effect. It is  assumed as usually  that interaction
between particles start at the time $t_{0}=-\infty$.

In order to obtain the Landau-Vlasov equation  the following
suggestions are adopted \cite{kmp92,kp94}: A) Born approximation
for two-body scattering; B) the time variation of the
nonequilibrium distribution function $\delta f \propto \exp
(-i\omega t)$ is taken periodic one with real frequency during all
interval of the time changing ($ - \infty \le t^{\prime} \le t $);
C) linear approximation on deviation of one-body Green functions
from their equilibrium values are used; D) Fermi system is
considered as  quasi-homogeneous one in coordinate space.

The linearized Vlasov-Landau equation has the form (\ref{LVone}) and
can be presented as \cite{kmp92,kp94}
\begin{equation}
\frac{d}{dt}f(\vec{p},\vec{r},t)+\{\varepsilon,f\}=
J(\vec{p},\vec{r},t) .
\end{equation}
Here,
\[
\{\varepsilon,f\} \equiv  \frac{\partial}{\partial \vec p}\varepsilon \cdot
\frac{\partial}{\partial \vec r}f - \frac{\partial}{\partial \vec
r}\varepsilon \cdot \frac{\partial}{\partial \vec p}f
\]
are the Poisson brackets and
\begin{equation}
\epsilon ( \vec{p}, \vec{r}, t) = \frac{p^2}{2\,m} + U (\vec {p}, \vec {r}, t)
\end{equation}
is the classical energy of a particle in the mean field
$U (\vec{p}, \vec{r}, t) =  \bar{\epsilon}+ \delta U (\vec{p}, \vec{r}, t)$,
where  expression  for $\delta U$  can be expressed in terms of the Landau interaction
amplitude $F(\vec p,\vec{p}^{\,\,'})$ for two-body collision scattering matrix:
\begin{equation}
\delta U = \frac{g}{N_F} \int \frac{d\vec{p}^{\,\,'}}{(2 \pi
{\hbar})^3}\,\, F(\vec p, \vec{p}^{\,\,'}) \,\, \delta f(
\vec{p}^{\,\,'},\vec r; t) ,
\label{h2}
\end{equation}
where $N_F = 2 \,p_F \,m^* /(g \,\pi^2 \,\hbar^3), \,\,\,p_F$ is
the Fermi momentum, $m^*$ is the effective mass of nucleon and $g$
is the spin degeneracy factor.

The linearized collision integral has form (\ref{eqJ})  with
(see, Eqs.(42),(43) and (45),(46) of Ref.\cite{kp94} for details)
\begin{equation}
J^{(j)}({\vec p}, {\vec r}, t) = 2 \int \frac{d{\vec p}_2\,d{\vec p}_3
\,d{\vec p}_4}{ (2\,\pi\,\hbar)^6}\, W(\vec p_1,\vec p_2,\vec p_3,\vec p_4)
\delta (\Delta {\vec p}) \,B^{(j)} ({\vec p}, {\vec r}, t) .
\label{eqsJj}
\end{equation}
Here, $W(\{{\vec p}_i\})=(d\sigma/d\Omega) 4(2\pi\hbar)^3/m^2 $ is
the probability  of two-body collisions with the initial momenta
${\vec p}_{1}= {\vec p}, {\vec p}_{2}$ and final ones $ {\vec
p}_{3}, {\vec p}_{4}$;  $d\sigma/d\Omega$ is in-medium
differential cross-section (in Born approximation);
\begin{equation} 
\label{eqsb12}
\begin{array}{l}
{\displaystyle B^{(1)}({\vec p}, {\vec r}, t) = \sum_{k=1}^4
\,\,\delta f_k(t) \, \frac{\partial \,Q(\{\bar{f}_j\})}{\partial
\,\bar{f}_k} \, [\delta_{+}(\Delta \bar{\epsilon} + \hbar
\,\omega) + \delta_{-}(\Delta \bar{\epsilon} - \hbar \,\omega)],
}\\{\displaystyle B^{(2)}({\vec p}, {\vec r}, t) {=}
Q(\{\bar{f}_j\}) \frac{\Delta (\delta U(t))}{\hbar\,\omega}
\{[\delta_{+}(\Delta \bar{\epsilon} {+} \hbar \omega) {-}
\delta_{+}(\Delta \bar{\epsilon})] {-} [\delta_{-}(\Delta
\bar{\epsilon} {-} \hbar \omega) {-} \delta_{-}(\Delta
\bar{\epsilon})]\}},
\end{array}
\end{equation}
 where $\bar{f}_{k} \equiv \bar{f}({\vec p}_{k},{\vec r} )$;
$\partial Q(\{\bar{f}_j\})/\partial \bar{f}_k$
are the derivatives of the is the Pauli blocking factor $Q$ with
respect to  $ \bar{f}_k$,
\begin{equation}
Q(\{\bar{f}_j\}) = (1-{\bar f}_{1})(1-{\bar f}_{2}){\bar f}_{3}{\bar f}_{4} -
{\bar f}_{1}{\bar f}_{2}(1-{\bar f}_{3})(1-{\bar f}_{4}) .
\label{Pauli}
\end{equation}
The ${\bar \epsilon}_{i}= \bar{\epsilon}(\vec{p}_{i}, {\vec r})$
and $\delta U_{j}$ are the equilibrium single-particle energy
and the variation of the mean field for particle with momentum
$\vec{p}_{i}$ respectively;
$
\Delta \bar{\epsilon} = \bar{\epsilon}_{1} + \bar{\epsilon}_{2} -
\bar{\epsilon}_{3} - \bar{\epsilon}_{4},
$
$
\Delta (\delta U) \equiv \delta U_1 + \delta U_2 -
\delta U_3 - \delta U_4,
$
$ \Delta {\vec p}= {\vec p}_{1}+ {\vec p}_{2}-{\vec p}_{3}-
{\vec p}_{4}.
$
The  equilibrium distribution function $ \bar{f}_{k} \equiv
\bar{f} ({\vec p}_{k},{\vec r})$ depends on the equilibrium
single-particle energy  $\bar{\epsilon}_{k} \equiv
\bar{\epsilon}({\vec p}_{k},{\vec r})$: $\bar{f}_{k} = {\bar f}
(\bar{\epsilon}_{k})$. It equals the Fermi function evaluated at
the temperature $T$, $\bar{f}(\bar{\epsilon}_{k}) =
1/[ 1 + \exp (({ \bar{\epsilon}_{k} - \mu) / T })]$.

Note that the generalized functions $\delta_{+}$, $\delta_{-}$ appearing in
Eq. (\ref{eqsb12})
include also integral contribution,
\begin{equation} 
\delta_{+}(x) = \frac 1{2 \pi}\,\int_{-\infty}^0 d\tau \,\,
e^{-i\,x\,\tau} = \frac{i}{2 \pi} \,\frac{1}{x + i0} = \frac 12
\delta(x) - \frac{1}{2  \pi i} {\cal P}(\frac{1}{x}) , \ \ \
\delta_{-}(x)=\delta_{+}^{*}(x) ,
\label{genfun}
\end{equation}
where $\delta(x)$ is the delta function and the symbol ${\cal P}$
denotes the principal value of integral contribution. The integral
terms of the  $\delta_{\pm}$, corresponding to virtual
transitions, are usually rejected in the $J$ because they assumed
to be included by renormalizing the interactions between particles
\cite{ryt86}. This corresponds to substitution of the $\delta
(x)/2$ for $\delta_{\pm}$ in Eq. (\ref{eqsb12}), i.e., to taking
into account real transitions with conservation of energy. We will
consider below only these transitions.

Now we will modify the expression for quantity $B^{(1)}$. We present
the nonequilibrium component $\delta f$ of the distribution function
in the form
\begin{equation}
\delta f(\vec {p}_j, \vec {r},t) = \, - \, \nu(\vec{p}_j, \vec {r},t)
\frac{ \partial \bar{f}(\bar{\epsilon}_j)}{ \partial \bar{\epsilon}_j} .
\label{df}
\end{equation}
Then the quantity $B^{(1)}$ can be written as
\begin{equation} \label{b1trans} 
\begin{array}{rcl}
{\displaystyle B^{(1)}({\vec p}, {\vec r}, t) }&=&{\displaystyle
\, - {1 \over 2}\, \sum_{k=1}^4 \,\,\nu_k \, \frac{\partial
\,Q(\{\bar{f}_j\})}{\partial \,\bar{\epsilon}_k} \,
[\delta(\Delta \bar{\epsilon} + \hbar \,\omega) +
\delta(\Delta \bar{\epsilon} - \hbar \,\omega)]}
\\&=&{\displaystyle {1 \over 2} \Delta \nu \,Q(\{\bar{f}_j\}) \frac{\partial \,}
{\partial \,\hbar \omega} \, [\delta(\Delta \bar{\epsilon} +
\hbar \,\omega) + \delta(\Delta \bar{\epsilon} - \hbar
\,\omega)] - \delta B^{(1)} ,}
\end{array}
\end{equation}
where $ \Delta \nu \equiv \nu_1 + \nu_2 -\nu_3 - \nu_4$, $\nu_{k}
= \nu (\vec{p}_{k}, \vec{r},t)$ and
\begin{equation}
\label{b1corr} 
\begin{array}{rcl}
\delta B^{(1)} &=& {\displaystyle {1 \over 2}\sum_{k=1}^4 \,
\frac{\partial}{\partial \,\bar{\epsilon}_k} \, \{ \nu_k
Q(\{\bar{f}_j\}) [\delta (\Delta \bar{\epsilon} + \hbar
\,\omega) + \delta(\Delta \bar{\epsilon} - \hbar \,\omega)]\}
} \\ &+& {\displaystyle {1 \over 2} \sum_{k=1}^4 \,Q(\{\bar{f}_j\})
[\delta (\Delta \bar{\epsilon} + \hbar \,\omega) +
\delta(\Delta \bar{\epsilon} - \hbar \,\omega)]
\frac{\partial \nu_k }{\partial \,\bar{\epsilon}_k} .}
\end{array}
\end{equation}

The first component in the Eq.(\ref{b1corr}) determines a
probability flux of colliding particles which is connected with
possibility of variation of the energy $\bar{\epsilon}_{k}$ when
the values of other energies ($\bar{\epsilon}_{j \neq k}$ and
$\hbar \omega$) are fixed. This term should be equal zero because
of fixing the total energy in initial or final states and
therefore it does not contribute to the total number of the
collisions ${\cal N}$:
\begin{equation} {\cal N}(\hat{p})
\equiv \int^{\infty}_{0} d\epsilon J( \hat{p},\epsilon) , \,\,\,\,
\hat{p} \equiv \vec {p}/ p .
\label{number}
\end{equation}

A relative dynamical component $\nu_{k}$ of the distribution
function  is slowly dependent on energy and it can be considered
(at least for low temperatures $T \ll \epsilon_{F}$) as a function
of the momentum direction rather than of the momentum: $\nu_{k}
\equiv \nu ( \vec {p}_{k}, \vec {r},t) = \nu ( \hat{p}_{k},\epsilon_F, \vec {r},t)$.
Therefore the second component in the Eq.(\ref{b1corr}) is also negligible
and the term $\delta B^{(1)}$  should be omitted from the Eq.(\ref{b1trans}),
$\delta B^{(1)} = 0 $. The expressions for $B^{(j)}$ take the form
\begin{equation} 
\label{eqsb12f}
\begin{array}{l}
{\displaystyle B^{(1)}({\vec p}, {\vec r}, t) = {1\over 2}
 \Delta \nu \,Q(\{\bar{f}_j\}) \frac{\partial \,}
{\partial \,\hbar \omega} \, [\delta(\Delta \bar{\epsilon} +
\hbar \,\omega) + \delta(\Delta \bar{\epsilon} - \hbar
\,\omega)]  ,}
\vspace{2mm} \\
{\displaystyle B^{(2)}({\vec p}, {\vec r}, t) = {1\over 2}
Q(\{\bar{f}_j\}) \frac{\Delta (\delta U(t))}{\hbar\,\omega}
\{[\delta(\Delta \bar{\epsilon} {+} \hbar \omega) {-}
\delta(\Delta \bar{\epsilon})] {-} [\delta(\Delta
\bar{\epsilon} {-} \hbar \omega) {-} \delta(\Delta
\bar{\epsilon})]\}} .
\end{array}
\end{equation}

The shift in energy $\Delta \bar{\epsilon}$ by  $\hbar \,\omega$ in the
arguments of the $\delta$-functions of the expressions (\ref{eqJ}), ({\ref{eqsJj}),
(\ref{eqsb12f}) for the collision
integral agrees with the interpretation of the collisions in the presence
of the collective excitations proposed by Landau \cite{LAND57}. According to
this interpretation, high-frequency oscillations in Fermi liquid can be
considered as phonons, that are absorbed and created at the two-particle
collisions.

As it was discussed in  the introduction, the form of  expressions
\ref{eqJ}), ({\ref{eqsJj}), (\ref{eqsb12f}) for
collision integral is not correct in general case and  an additional modification
of  these expressions are needed. The incorrectness results from
approximations  which were made at the kinetic equation derivation.

Because of assumptions both on undamped behaviour of the distribution function
throughout all time interval ($ - \infty \le t^{\prime} \le t \to \infty$)
and on small magnitude of two-body interaction,
one can expect overestimation of retardation effects in the foregoing
expression for collision integral. It means that from physical point of
view this relationship should be  fulfilled only in the case of small $\omega$.

Therefore, we replace the derivatives of the form
$\partial \, \delta(\Delta \bar{\epsilon} + \hbar \,\omega) /
\partial \,\hbar \omega$ and
$\partial \, \delta(\Delta \bar{\epsilon} - \hbar \,\omega) /
\partial \,\hbar \omega$ in $B^{(1)}$ of the Eq.(\ref{eqsb12f}) by the finite
differences $(\delta(\Delta \bar{\epsilon} + \hbar \,\omega) -
\delta(\Delta \bar{\epsilon})) /  \hbar \,\omega$ and
$(\delta(\Delta \bar{\epsilon}) -
\delta(\Delta \bar{\epsilon} - \hbar \,\omega))  /  \hbar \,\omega$,
respectively. Then we combine the resulting expression together with
contribution $B^{(2)}$ arising from mean-field variation  and
finally obtain the linearized collision integral for  Fermi
liquid in the following form
\begin{equation} 
J({\vec p},{\vec r},t) =
 \int \frac{d{\vec p}_2 d{\vec p}_3 d{\vec p}_4}{ (2\pi\hbar)^6}
  W(\{{\vec p}_i\}) \delta (\Delta {\vec p}) \Delta \chi Q
\frac{\delta (\Delta \epsilon +\hbar\omega) - \delta ( \Delta
\epsilon -\hbar\omega)}{\hbar \omega } .
\label{int1}
\end{equation}

Here, $ \Delta \chi \equiv \chi_1 + \chi_2 -\chi_3 - \chi_4$;
$\chi_{k}= \chi ( \vec{p}_{k}, \vec {r},t)$ is a function
determining relative deviation  of distribution function
from local equilibrium state, $\delta \bar{f}$, (\ref{flocal}):
\begin{equation}
\delta \bar{f}= \delta f -
\frac{d \bar{f}}{d \bar{\epsilon}} \delta U =
f( \vec {p}, \vec{r},t) -f_{l.e.}
= \, - \, \chi \frac{d \bar{f}}{d \bar{\epsilon}},
 \ \ \chi = \nu + \delta U .
\label{relflocal}
\end{equation}
With the use of the algebraic relation \cite{BP1991}
\begin{equation} 
[(1-{\bar f}_{1})(1-{\bar f}_{2}){\bar f}_{3}{\bar f}_{4} - {\bar
f}_{1}{\bar f}_{2}(1-{\bar f}_{3})(1-{\bar f}_{4}) \exp
\left(\frac{\mp \hbar \omega}{T} \right)] \delta (\Delta \epsilon
\pm \hbar \omega)=0 ,
\end{equation}
the Eq. (\ref{int1})  can be presented as
\begin{equation} \label{int2} 
\begin{array}{l}
J(\vec p,\vec r,t)=\vspace{2mm} \\{\displaystyle \int\frac{d\vec
p_2 d\vec p_3 d\vec p_4}{(2\pi\hbar)^6} W(\{\vec
p_i\})\delta(\Delta \vec p)\Delta\chi {\bar f}_{1}{\bar
f}_{2}(1-{\bar f}_{3})(1-{\bar f}_{4})
\left[\Phi(\hbar\omega,T)-\Phi(-\hbar\omega,T)\right] ,}
\end{array}
\end{equation}
\noindent where $\Phi(\hbar \omega, T) = \delta(\Delta\epsilon+\hbar\omega)
[\exp(-\hbar \omega /T) -1]/ 2 \hbar \omega$.

The collision integral  of the form (\ref{int1}) or (\ref{int2})
depends on the variation $\delta \bar {f}$, $J \equiv  J(\delta \bar{f})$.
It was mentioned in the introduction this behaviour is in line with
general properties of the Vlasov-Landau equation in the
Fermi-liquid \cite{BP1991,LifPit} at $\partial \delta f / \partial
t = 0$. It provides driving distribution function towards its local
equilibrium value. This behaviour is in line with
general properties of the Vlasov-Landau equation in the
Fermi-liquid \cite{BP1991,LifPit} at $\partial \delta f / \partial
t = 0$.
The expressions (\ref{int1}), (\ref{int2}) depend only on
the occupation probability  ${\cal P}_{2p2h}\equiv {\bar
f}_{1}{\bar f}_{2}(1-{\bar f}_{3})(1-{\bar f}_{4})$ of the 2p-2h
states in the phase space. This fact leads to interpretation of
the collision damping with linearized collision term as the
relaxation process due to the coupling of one-particle and
one-hole states to more complicated $2p-2h$ configurations.

The form of the collision integral (\ref{int2}) in the Markovian limit
($\omega \rightarrow 0$) coincides with the standard expression for
the collision integral in Fermi-liquid without retardation
effects~\cite{BP1991,LifPit} because in this case the term in square
brackets of Eq.(\ref{int2}) tends to the value
$- \delta(\Delta \epsilon) / T$.

The equation (\ref{int1}) for some special explicit form of the quantity
$\chi_{j}$ was used at first in Refs.~\cite{AB92,abb95,bab95}. The derivation
of the collision integral (\ref{int1}) is performed in Ref.~\cite{AS98} within
framework of the extended time-dependent Hartree-Fock model.
The expressions for the distortion functions $\chi_{j}$  corresponding to
a perturbation approach on collision term and including the
amplitudes of the random phase approximation were used in this method.

Note that the expression for the collision integral in two-component Fermi-system
is obtained from Eq.(\ref{int2}) in the same manner as done in Ref.\cite{PLU99}
under the assumption that chemical potentials and  the equilibrium
distribution functions  are  the same for protons and neutrons.

\section{Calculations of the relaxation times and nuclear matter viscosity}

The collision integral can be used to calculate collisional relaxation times
governing  the dissipative properties of different physical quantities
\cite{SB1970,BP1991,AB92,KMPNP1992,kmp92,kmpzp2,AKoeh93,Bertsch78,KPS95}.
Below we calculate relaxation times, $\tau^{(\pm)}_{\ell}$, of collective vibrations in
two-component nuclear matter consisting of neutron and protons subsystems.
These  collective relaxation times is  determined by interparticle collisions
within the distorted layers of the Fermi surface with multipolarity $\ell$:
\begin{equation} 
{1\over \tau^{(\pm)}_{\ell }} \equiv  \int^{\infty }_{0}
d\epsilon_1 \int d\Omega_p J^{(\pm)}_c(\hat{p},\epsilon_{1})
Y_{\ell 0}(\hat{p})/
\int^{\infty }_{0}d\epsilon_{1} \int d\Omega_{p}
\delta f^{(\pm)} Y_{\ell 0}(\hat{p}) ,
\label{tau}
\end{equation}
where $Y_{\ell m}(\hat{p})$ is the spherical harmonic function and
$J^{(\pm)}(\hat{p},\epsilon)$ are the linear combinations of the  collision
integrals for protons $J_{p}$ and neutrons $J_{n}$ in nuclear matter:
$ J^{(\pm)}  = (J_{p} \pm J_{n})/2 $; $\delta f^{(\pm)} =
(\delta f_{p} \pm \delta f_{n})/2$.
These times  are
proportional to the relaxation times $\tau^{(\pm)}_{c}$ defining the
damping widths $\Gamma^{(\pm)}_{c}(L)$ of the isoscalar $(+)$ and the isovector
$(-)$ vibrations with frequency $\omega$ \cite{kmpzp2,PLU99,KPS95,KPS96} in
regime of rare collisions with $\omega \tau^{(\pm)}_{c} \gg 1$ in the
Fermi liquid. In particular, the collisional damping widths of giant
resonances with dipole ($L=1$) and quadrupole ($L=2$) multipolarities
resemble the widths in the relaxation rate approach
\begin{equation} 
\Gamma^{(\pm)}_{c}(L)=\hbar/\tau^{(\pm)}_{c}(L) , \ \ \
\tau^{(-)}_{c}(L=1) = \tau^{(-)}_{\ell=1}, \ \ \
\tau^{(+)}_{c}(L=2) = \tau^{(+)}_{\ell=2} ,
\label{width}
\end{equation}
in the case when nuclear fluid dynamical model with  relaxation is used
\cite{PLU99,KPS95}. The collisional damping
width\cite{kmpzp2} of zero sound in the Fermi liquid with its relative velocity
$S_{r}\simeq 1$ is also given by Eq.(\ref{width}) but with the use of the
$\tau^{(+)}_{\ell \to \infty} \propto \tau^{(+)}_{2}$ for
$\tau^{(\pm)}_{c}(L)$.
The time $\tau^{(+)}_{\ell=2}$ at $\omega=0$
is the thermal relaxation time determining the viscosity
coefficient of the Fermi liquid~\cite{Bertsch78}.

The time $\tau^{(+)}_{\ell=2}$ at $\omega=0$
is the thermal relaxation time determining the viscosity
coefficient of the Fermi liquid.

The analytical expressions for collective relaxation times
of the damping of the collective vibrations with frequency  $ \omega$
has the following general form in low-temperature  and low-frequency
limits ($T, \hbar \omega\ll \epsilon_F$)\cite{kmpzp2,PLU99,PGK2001}
\begin{equation}
{\hbar \over \tau^{(\pm)}_{\ell}} =
{ 1 \over \alpha^{(\pm)}_{\ell} }
\left\{(\hbar \omega)^2+(2\pi T)^2 \right\} , \ \ \
{ 1 \over \alpha^{(\pm)}_{\ell} }={ 2 m\over 3\pi \hbar^{2}}
\left[ <\sigma^{\prime}_{av} \Phi^{(+)}_{\ell }>+
< \sigma^{\prime}_{pn} \Phi^{(\pm)}_{\ell }> \right] ,
\label{atau}
\end{equation}
\noindent where $ \sigma^{\prime}_{av}=
( \sigma^{\prime}_{nn}+\sigma^{\prime}_{pp} )/2$;
$\sigma^{\prime}_{j j^{\prime}} \equiv d\sigma_{j j^{\prime}}/d\Omega$ is
in-medium differential cross-section for  scattering of the nucleons $j$ and
$j^{\prime}$ ( here, $j=n$ or $p$, and similarly $j^{\prime}= p$ or $n$).
The symbol $<\ldots >$ in Eq.(\ref{atau}) denotes averaging over angles
of the relative momenta of the colliding particles,
\begin{equation}
<(\ldots)> = {1\over \pi} \int^{\pi}_{0}
d \phi \sin(\phi /2) \int^{\pi}_{0} d \theta (\ldots) .
\label{angint}
\end{equation}
The functions $\Phi^{(\pm)}_{\ell}$ define the angular
constraint on nucleon scattering within the distorted layers of the
Fermi surface with multipolarity $\ell$:
\begin{equation}
\Phi^{(\pm)}_{\ell } = 1 \pm P_{\ell }(\cos {\phi}) -
P_{\ell }((\hat{p}_{3} \hat{p}_{1})) \mp
P_{\ell }((\hat{p}_{4} \hat{p}_{1})) ,
\label{phi}
\end{equation}
where the scalar products $(\hat{p}_{3} \hat{p}_{1})$ and
$(\hat{p}_{4}\hat{p}_{1})$ are given as
\begin{eqnarray}
(\hat{p}_{3} \hat{p}_{1}) &=&
\cos^{2} (\phi/2) + \sin^{2} (\phi/2) \cos \theta ,
\nonumber \\
(\hat{p}_{4} \hat{p}_{1}) &=&
\cos^{2} (\phi/2) - \sin^{2} (\phi/2) \cos \theta .
\label{addangles}
\end{eqnarray}

Due to the momentum conservation and conditions $p_{i}$= $p_{F}$,
the angle $\theta$ agrees with the scattering angle in the center-of-mass
reference frame of two nucleons. The angle $\phi$ defines the magnitudes
of the relative momenta $ \vec{k} _{i} = (\vec{p}_{2}-\vec{p}_{1})/2$
and  $ \vec{k}_{f} = (\vec{p}_{4}-\vec{p}_{3})/2$ before and after collision,
respectively. The value of total momentum,
$\vec{P}=\vec{p}_{1}+\vec{p}_{2}$, also depends on a
magnitude of the $\phi$. We have
\begin{equation}
\vec{k}_{i} \vec{k}_{f} = k^{2} \cos \theta , \ \
k^{2} = k^{2}_{i} = k^{2}_{f} = p^{2}_{F} \sin^{2}(\phi/2) , \ \
\vec{P}^{2} = 4 p^{2}_{F} \cos^{2}(\phi/2) .
\label{reldef}
\end{equation}
Therefore the relative kinetic energy $E_{rel}$ of two nucleons as well as
the energy of centrum mass motion $E_{cm}$ are dependent on angle $\phi$
\begin{equation}
E_{rel} = k^{2}/m = 2 \epsilon_{F} \sin^{2}(\phi/2) , \ \ \
E_{cm} = P^{2}/2m = 2 \epsilon_{F} \cos^{2}(\phi/2)
\label{endef}
\end{equation}
and the total energy $E_{tot}=E_{rel}+E_{cm}$  holds only fixed,
$E_{tot}= 2 \epsilon_{F}$. Therefore the in-medium differential
cross-sections $\sigma^{\prime}_{j,m}$ of two nucleon scattering depend
on the relative momenta $\vec{k}_{i}$ and $\vec{k}_{f}$ at fixed total energy
rather then at fixed relative kinetic energy $E_{rel}$, because the magnitude
of $E_{rel}$ changes with angle $\phi$ between colliding particles.
The transfer momenta
$\vec{q} = \vec{k}_{i} -\vec{k}_{f} = \vec{p}_{3} -\vec{p}_{1}$ and
$\vec{q}^{\prime} = -( \vec{k}_{i} +\vec{k}_{f}) = \vec{p}_{1} -\vec{p}_{4}$
for scattering due to direct and exchange interactions respectively are
also functions of the $\phi$ and $\theta$:
$q = 2 k(\phi) \sin (\theta/2)$ and $q^{\prime} = 2 k(\phi) \cos (\theta/2)$.

Now we estimate the collisional relaxation times in the case of the
isotropic scattering with independent of energy the angle-integrated cross
sections $\sigma_{j j^{\prime}}$.
Performing angular integration in (\ref{atau}) with the use of
Eqs.(\ref{angint}) and (\ref{phi}) we find that
$1 / \tau^{(\pm)}_{\ell < \ell^{(\pm)}_{0}} = 0$   and
\begin{equation} 
{\hbar \over \tau^{(\pm)}_{\ell}} =
{ 1 \over \alpha^{(\pm)}_{\ell} }
\left[(\hbar \omega/2 \pi)^2+T^2 \right] , \ \ \
{ 1 \over \alpha^{(\pm)}_{\ell} }={ 8 m\over 3 \hbar^{2}}
\left[ c_{\ell} \sigma_{av} + d^{(\pm)}_{\ell} \sigma_{np} \right] ,
\label{taucon}
\end{equation}
$$
c_{\ell}=1-{ 2-(-1)^{\ell} \over 2\ell+1} , \ \
d^{(-)}_{\ell}={1-(-1)^{\ell}\over 2\ell+1} , \ \
d^{(+)}_{\ell}=d^{(-)}_{\ell=0}=c_{\ell=0}=c_{\ell=1}=0 ,
$$
where $\sigma_{av}$ = $[\sigma_{pp} + \sigma_{nn} +2 \sigma_{np}]/4$ is
the in-medium spin-isospin averaged nucleon-nucleon cross section.
The magnitude of the in-medium  cross section $\sigma_{j j^{\prime}}$
is taken  usually proportional to the value of the
free space cross section $\sigma^{(f)}_{j j^{\prime}}$ with a factor
$F= \sigma_{j j^{\prime}} / \sigma^{(f)} _{j j^{\prime}}$, so that
the parameter $\alpha^{(\pm)}_{\ell}$ can be rewritten in the form
\begin{equation}
\alpha^{(\pm)}_{\ell}= \widetilde{\alpha}^{(\pm)}_{\ell} /F, \ \
\widetilde{\alpha}^{(\pm)}_{\ell}=
4.18 / \left[ c_{\ell} + 1.3 d^{(\pm)}_{\ell}\right], \ \ MeV.
\label{falpha}
\end{equation}

Here, the values $\sigma^{(f)}_{av} =3.75~fm^2$ and
$\sigma^{(f)}_{np} = 5~fm^2$ are adopted \cite{AB92,KPS96}; they
correspond to the free space cross sections near Fermi energy. The
dependence on $\ell$ for relaxation times $\tau^{(\pm)}_{\ell}
\equiv \tau^{(\pm)}_{\ell,f}$ with the use of free space cross
sections is shown in Fig. 1; $\hbar\omega =13.43 MeV$. The Figs.
2,3 describe relative relaxation times versus multipolarity $\ell$
for cross section parameterizations using Gogny and Skyrme
effective two-body forces with parameters from
\cite{AS98,YGYA2000}. The relative relaxation times presented in
Fig. 3 were calculated using cross section on Fermi surface. Solid
and dashed lines connect the values which correspond to isoscalar
and isovector modes of vibrations respectively.

The magnitudes of the relaxation times are different for isoscalar
and isovector modes of vibrations  and they are dependent on the
multipolarity $\ell$.  The collisional relaxation times  rather
slowly vary  with collective motion mode at isotropic scattering
with energy independent free cross sections. As seen from the Fig.
3, the relaxation times calculated with effective interaction
between nucleons are greater than that ones with cross-section in
free space. It means that in-medium cross-section between nucleons
in nuclear matter is smallest than in free space ($\approx $ 20~\%
for Skyrme forces and $\approx $ 60~\% for Gogny interaction). The
relaxation times $\tau^{(\pm)}_{\ell}$ depend on  frequency
$\omega$ due to the memory effects in the collision integral. The
temperature dependence arises from smearing out the equilibrium
distribution function near the Fermi momentum in  heated nuclei.
The collisional rates $1 / \tau^{(\pm)}_{\ell}$ are quadratic both
in temperature and in frequency with the same relationship between
the components much as in the zero sound attenuation factor of
heated Fermi liquid within the Landau prescription
\cite{LAND57,KPS95,AB92,AS98}.

According to response function approach the damping width of giant isovector
dipole resonance (GDR)  at the temperature $T$  is presented
in the form \cite{PEK2001}
\begin{equation}
\Gamma(T)
= 2\,q  \gamma
{E^{2}_{r}+E^{2}_{0} \over
(E^{2}_{r}-E^{2}_{0})^{2} + (2 \gamma E_{r})^{2}} ,
\label{g6}
\end{equation}
\noindent where $\gamma$ is determined by  relaxation time
$\tau_{c}(\hbar\omega = E_r,T)$,
\begin{equation}
\gamma =  {\hbar \over \tau_{c}(\hbar\omega = E_r,T)},
\label{r48}
\end{equation}
\noindent  $E_r$ is an energy of giant dipole resonance; $E_{0}=
41 \,A^{-1/3}\,\, $ MeV. Here, we determine  quantity  $q$ from
equality of the GDR width in cold nuclei  with corresponding
experimental value $\Gamma_{exp}$: $ \Gamma(T=0)= \Gamma_{exp}$
The temperature dependence of the GDR width  according to
Eq.(\ref{r48}) in atomic nuclei ${}^{208}Pb$ and ${}^{120}Sn$ are
shown on Figs.6. Experimental data are taken from
Refs.\cite{HBH94,BR98} and they are indicated by points. The
relaxation time $\tau_{c}$ is taken as equal to
$\tau^{(-)}_{\ell=1}$: dash lines in Fig. 4 correspond to
calculations with Skyrme forces; dash-dot lines - Gogny
interaction; solid lines - calculations with free space
cross-section for two-nucleon collisions. The temperature
behaviour of the damping width is in rather close agreement with
the ones of experimental data.

It is seen from Figs. 4 that in rare collision regime  the
dependence of the GDR widths on the collective vibration frequency
and the temperature has the following  form $ \Gamma \propto
(\hbar \,\omega)^2 + 4\,(\pi \,T)^2$ , which corresponds to the
Landau's prescription \cite{LAND57,E1962,AIH83}.

The  relaxation times  rather slowly vary with multipolarity
of the Fermi surface distortions governed by collective motion and two-body
collisions. It gives possibility to use approximately the relaxation time ansatz
for collision integral.

This work is supported in part by the IAEA(Vienna) under contract
302-F4-UKR-11567.

\newpage
\pagestyle{empty}

\begin{figure}
\begin{center}
\hspace{-10mm}
 \includegraphics[width=0.8\textwidth,clip]{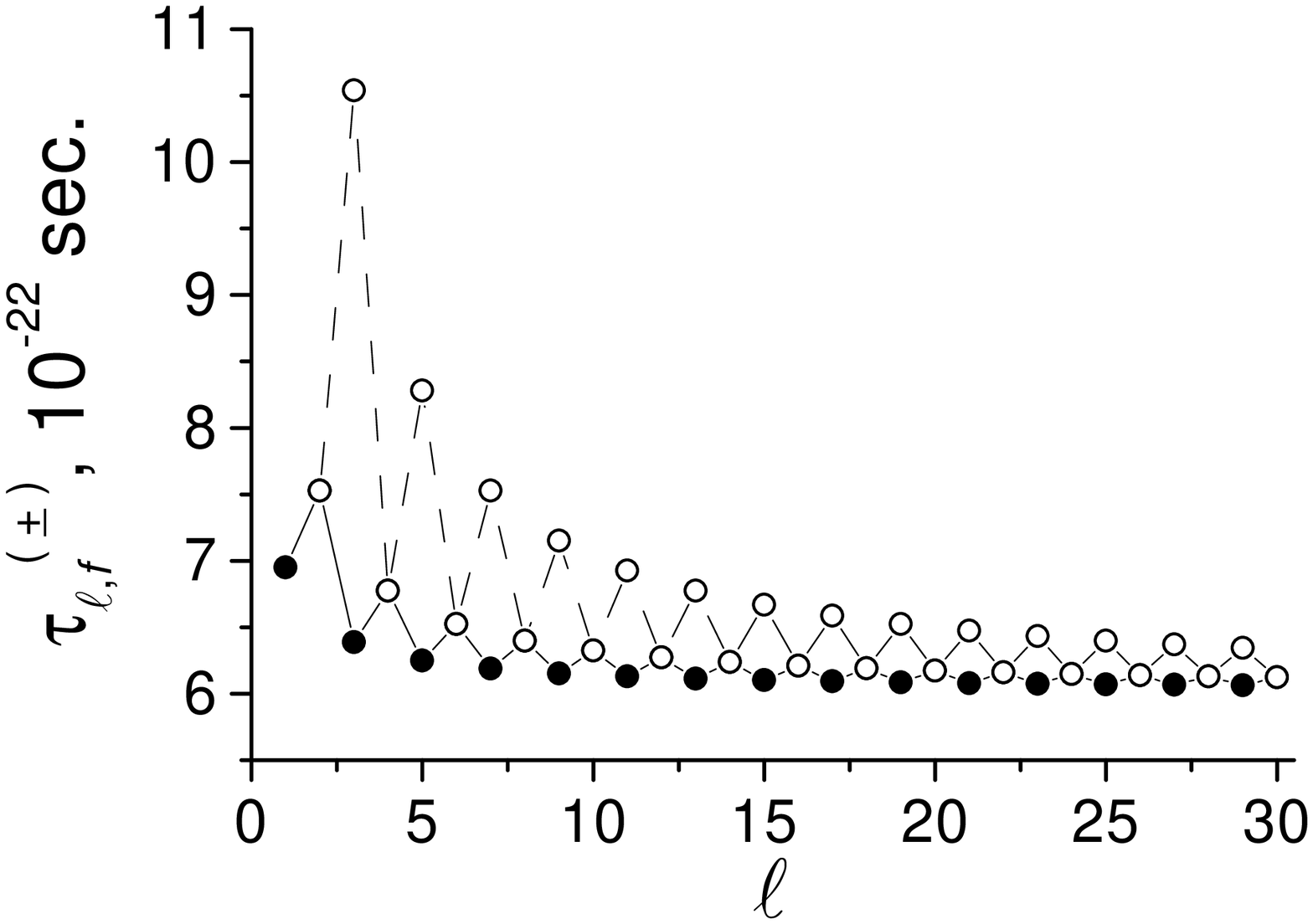}
\vspace{10mm}
 \parbox[t]{0.9\textwidth}{ Fig. 1: The relaxation times
 $\tau^{(\pm)}_{\ell,f}$
 versus multipolarity $\ell$ in cold nucleus $^{208}Pb$
  with free space cross section.}
 \vspace{1mm}
 \end{center}
\end{figure}

\begin{figure}
\begin{center}
\hspace{-10mm}
 \includegraphics[width=0.8\textwidth,clip]{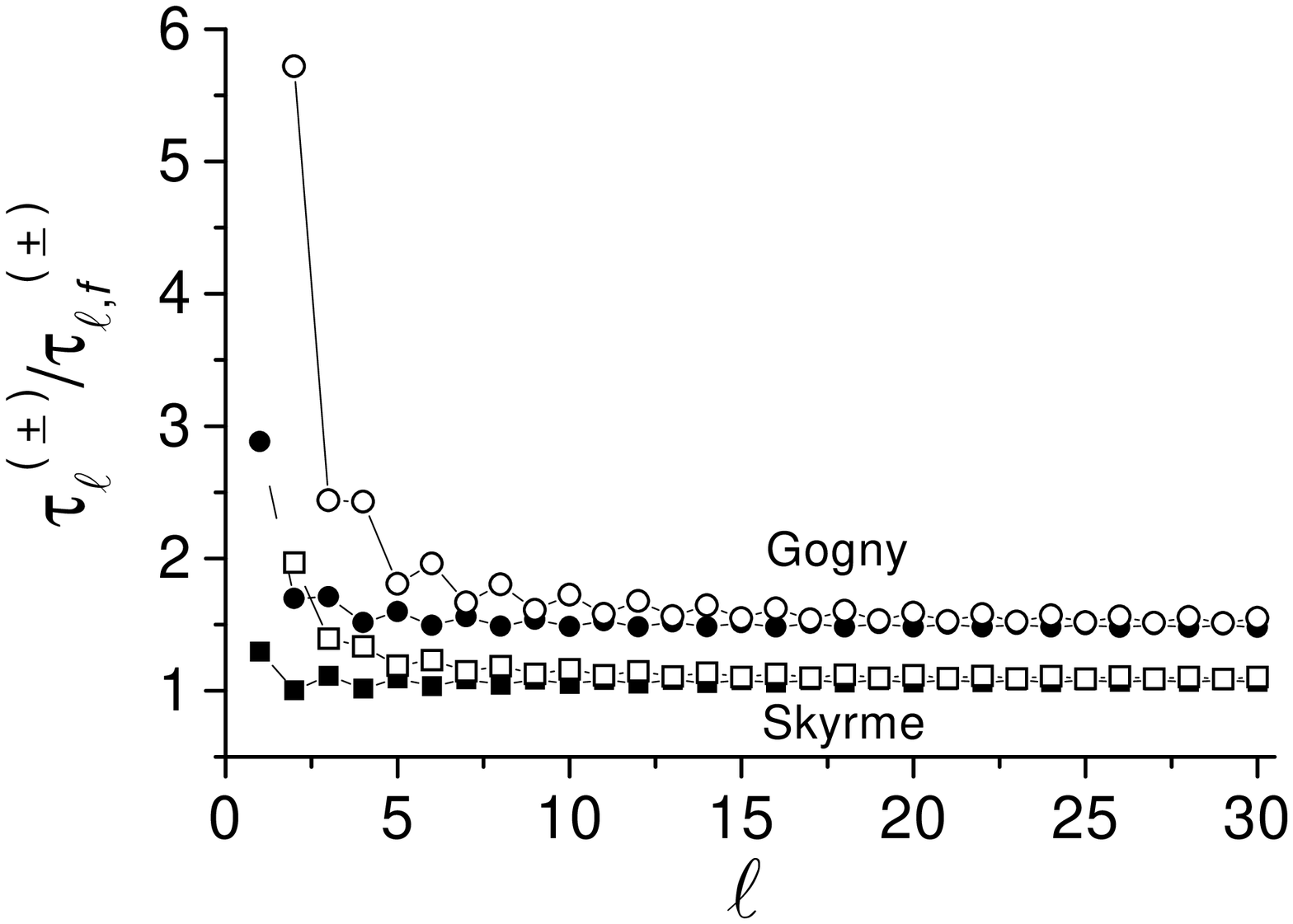}
\vspace{10mm}
 \parbox[t]{0.9\textwidth}{ Fig. 2: The relative relaxation times
   $\tau^{(\pm)}_{\ell}/\tau^{(\pm)}_{\ell,f}$ versus multipolarity
   $\ell$ for different cross section parameterizations.}
 \vspace{1mm}
 \end{center}
\end{figure}

\begin{figure}
\begin{center}
\hspace{-10mm}
 \includegraphics[width=0.8\textwidth,clip]{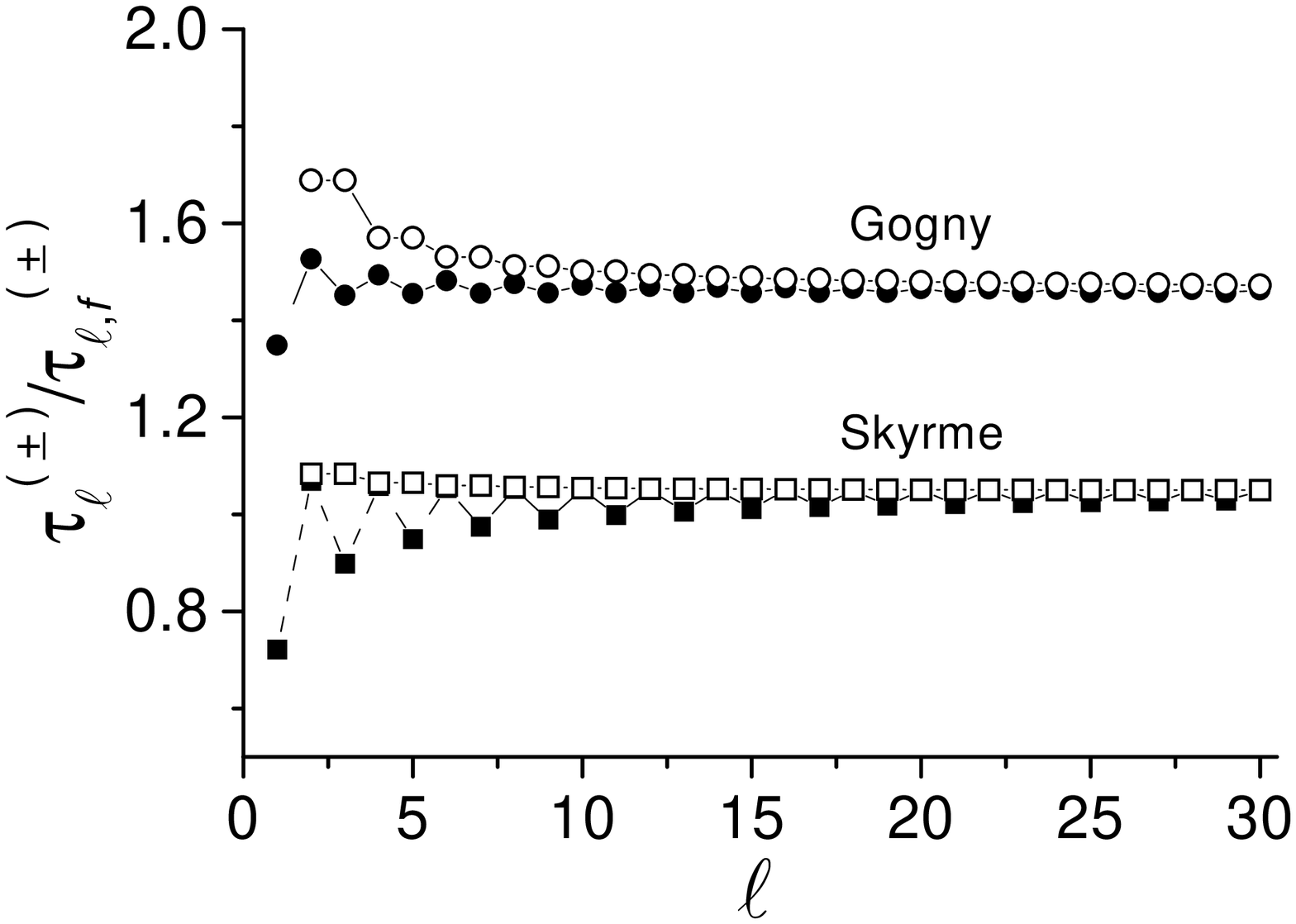}
\vspace{10mm}
 \parbox[t]{0.9\textwidth}{ Fig. 3: The relative
 relaxation times
   $\tau^{(\pm)}_{\ell}/\tau^{(\pm)}_{\ell,f}$ versus multipolarity
   $\ell$ for different cross section parameterizations calculated on
   Fermi surface.}
 \end{center}
\end{figure}

\begin{figure}
 \includegraphics[width=0.5\textwidth,clip]{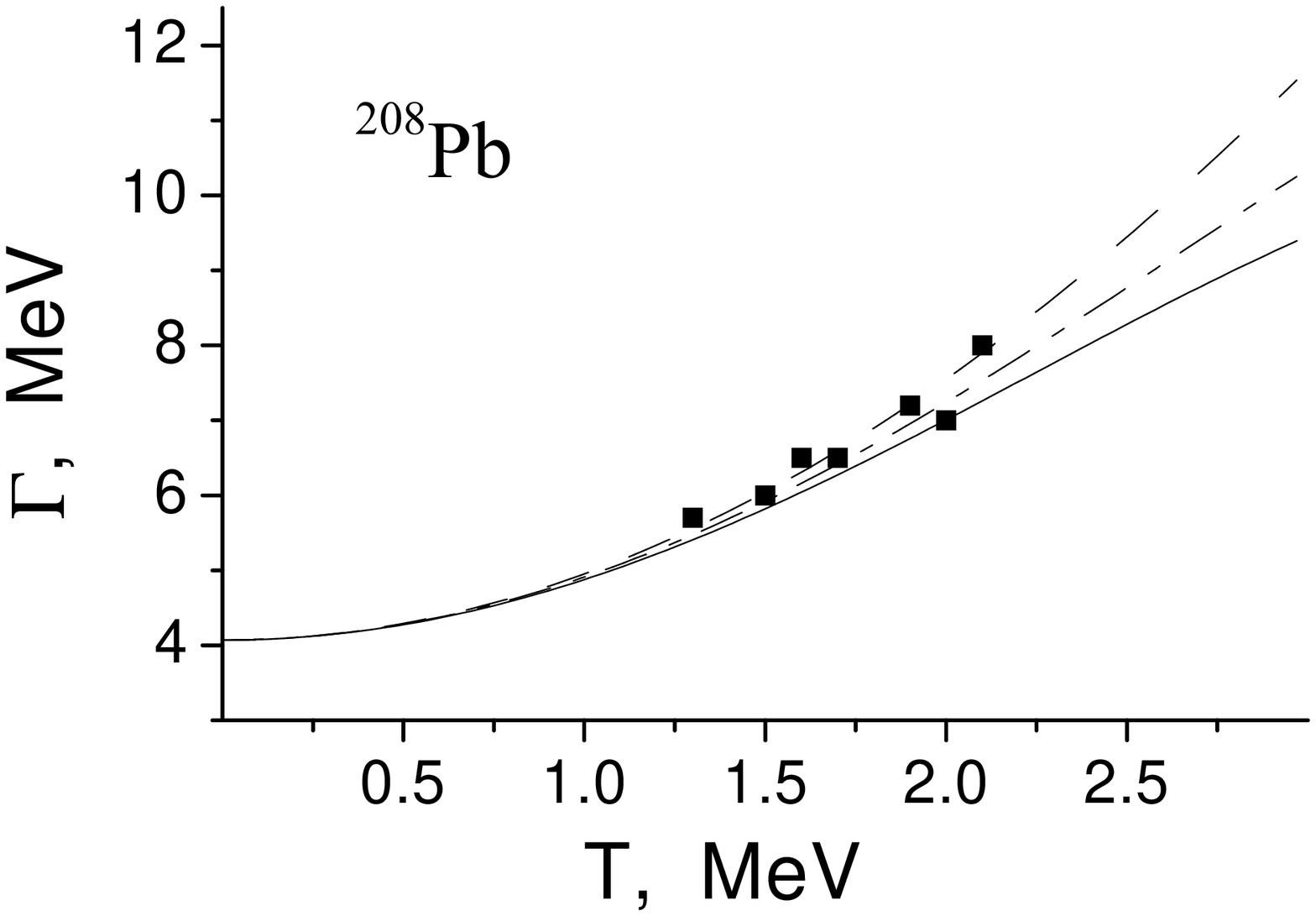}
 \hfill
 \includegraphics[width=0.5\textwidth,clip]{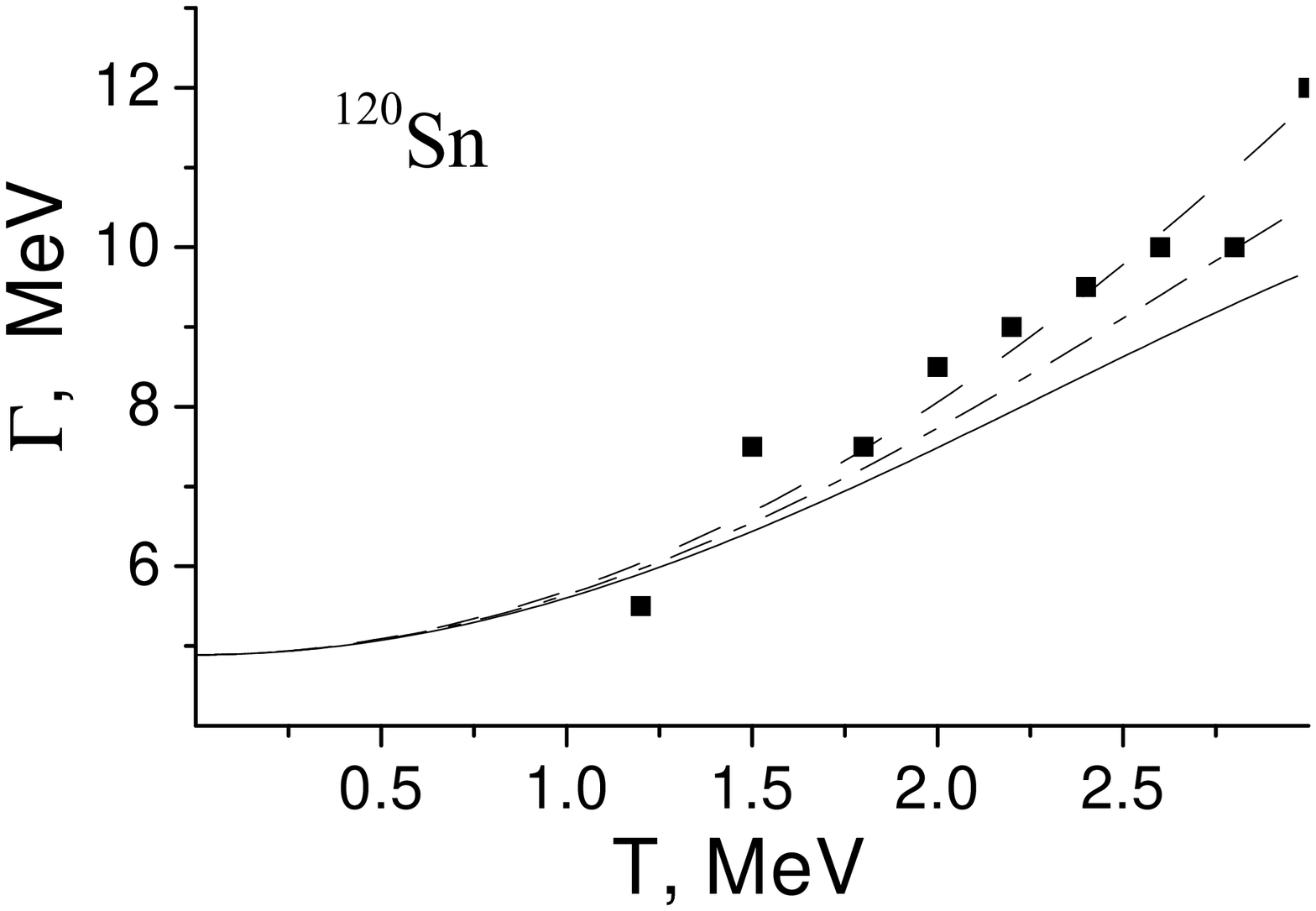}
 \parbox[t]{\textwidth}{Fig. 4:
 The temperature dependence of the GDR width  according to Eq.(\ref{r48})
in nuclei ${}^{208}Pb$ and ${}^{120}Sn$: dash lines - calculations with Skyrme forces;
dash-dot lines - Gogny interaction; solid lines - calculations with free space
cross-section for two-nucleon collisions. }
\end{figure}

\end{document}